\documentclass[12pt]{iopart}
\usepackage{graphicx}
\usepackage{amssymb}
\usepackage[english]{babel}
\usepackage{psfrag}
\newcommand{\bq}{\begin{equation}}
\newcommand{\eq}{\end{equation}}
\newcommand{\ba}{\begin{eqnarray}}
\newcommand{\ea}{\end{eqnarray}}

\usepackage{ifthen}

\newcommand{\pd}[2]{
	\ifthenelse{\equal{#2}{1}}{\frac{\partial}{\partial #1}}
	{\frac{\partial ^ #2}{\partial #1 ^#2}}
}
\newcommand{\dd}[2]{
	\ifthenelse{\equal{#2}{1}}{\frac{{\rm d}}{{\rm d} #1}}
	{\frac{{\rm d}^ #2}{{\rm d} #1 ^#2}}
}

\newcommand{\pf}[3]{
	\ifthenelse{\equal{#3}{1}}{\frac{\partial #1 }{\partial #2}}
	{\frac{\partial ^ #3 #1 }{\partial #2 ^#3}}
}
\newcommand{\df}[3]{
	\ifthenelse{\equal{#3}{1}}{\frac{{\rm d} #1}{{\rm d} #2}}
	{\frac{{\rm d} ^ #3 #1}{{\rm d} #2 ^#3}}
}

\begin{document}
\title{Mass condensation in one dimension with pair-factorized steady states}
\author{B. Waclaw$^1$, J. Sopik$^2$, W. Janke$^1$ and H. Meyer-Ortmanns$^2$}
\address{$^1$Institut f\"ur Theoretische Physik, Universit\"at Leipzig,\\Postfach 100\,920, 04009 Leipzig, Germany}
\address{$^2$School of Engineering and Science, Jacobs University,\\P. O. Box 750561, 28725 Bremen, Germany}
\eads{\mailto{Bartlomiej.Waclaw@itp.uni-leipzig.de}, \mailto{julien.sopik@googlemail.com}, \mailto{Wolfhard.Janke@itp.uni-leipzig.de}, \mailto{h.ortmanns@jacobs-university.de}}

\begin{abstract}
We consider stochastic rules of mass transport which lead to steady states that factorize over the links of a one-dimensional ring. Based on the knowledge of the steady states, we derive the onset of a phase transition from a liquid to a condensed phase that is characterized by the existence of a condensate.
For various types of weight functions which enter the hopping rates, we determine the shape of the condensate, its scaling with the system size, and the single-site mass distribution as characteristic static properties. As it turns out, the condensate's shape and its scaling are not universal, but depend on the competition between local and ultralocal interactions. So we can tune the shape from a delta-like envelope to a parabolic-like or a rectangular one. 
While we treat the liquid phase in the grand-canonical formalism, we develop a different analytical approach for the condensed phase. Its predictions are well confirmed by numerical simulations. Possible extensions to higher dimensions are indicated.
\end{abstract}
\maketitle

\section{Introduction}
Stochastic mass transport describes the movement of a generic mass from one space point to another as a classical stochastic process. ``Mass" stands for a generic quantity that is conserved throughout the process. It can be as different as a vehicle in macroscopic traffic or a macromolecule moving along the cytoskeleton \cite{chowdhury}. Above a certain density, mass condensation may be observed, so that even in the limit of infinitely many masses  a finite fraction of all masses piles up at a certain site, although the dynamic rules are fully symmetric. This is an expression of spontaneous symmetry breaking. In these systems it can happen even in one dimension, as they are in general out-of-equilibrium. When the mass transport describes vehicles, mass condensation corresponds to traffic jams. Other applications  of condensation are found in granular flow \cite{liu}, clustering \cite{majumdar}, they are manifest as gelation in networks where a single node takes a finite fraction of all links \cite{redner}, as Bose-Einstein condensation, or in phase transitions of quantum gravity \cite{grav}.

Many of these systems can be modeled as a set of particles occupying discrete levels, or ``boxes'' on a one-dimensional grid.
The balls-in-boxes model (B-in-B) \cite{bbj}, its non-equilibrium version, the zero-range process (ZRP) \cite{evans2000}, or a more general model \cite{evans2004} with continuous masses  are well-known examples.
Although these models are more abstract than realistic ones, they serve as paradigms since the stationary state can be derived analytically
as fully factorizing over the sites of the grid, or, more generally, over the nodes of an arbitrary graph \cite{2006}.
The factorization is due to ultra-local (``zero-range'') rules assumed to govern the dynamics of particles.
This makes the phase structure of the systems accessible and enables comparison with experiments on condensation phenomena, at least on a qualitative level.

As a natural consequence of the lack of interactions between particles at neighboring sites, the condensates in the B-in-B or ZRP models --- when they occur --- they always occupy a single site.
The question then arises as to how the shape of the condensate changes in the presence of interactions between sites that tend to flatten out the condensate's profile but still preserve translational symmetry and conserve the current. A first answer has been given in~\cite{evans1} for a model that is related to a solid-on-solid (SOS) model \cite{sos,sos2}, supplied with dynamical rules that drive the system out-of-equilibrium. There, the steady state factorizes over {\em pairs\/} of sites of a one-dimensional ring topology which allows for nearest-neighbour interactions while making the system analytically solvable. The condensate is extended over a range that scales as $\sqrt{N}$ if $N$ is the system size. The hopping rates were determined by weights $g(m,n)$ that factorize into an ultralocal part, depending on the occupation numbers of a single site, and a local part, depending on the occupation number of neighboring sites. For this model, the scaling of the condensate's extension $W$ with $\sqrt{N}$ can be derived as the first-return time of a random walker after $W$ time steps (see \cite{evans1}). This is possible when the very formation of the condensate is described in terms of a random walk, for which the walker chooses his step sizes according to a distribution that is given by the weights $g(m,n)$. One may naively expect that this analogy carries over also to other choices of weight functions, so that an extension with $\sqrt{N}$ would be a universal feature of condensates with PFSS.

In this paper we shall show that this is not the case, i.e., the scaling is non-universal. Instead we can actually tune the shape of the condensate and its extension via different appropriately chosen weight functions. In terms of a random walk, the walker would no longer use the weights $g(m,n)$ as distribution that determines his choice of step sizes, so that we have to look for another derivation.
As it turns out, the relevant characteristics for both local and ultralocal weight factors is the range of interactions in occupation-number space: we call short-range interactions local weight factors $K(x)=K(\vert m-n\vert)$ and ultralocal factors $p(m)$ which decay faster than any power in their argument, and long-range interactions those which decay as a power of their argument ($x$ or $m$, respectively).
For local interactions with exponential decay (short-range) and an ultralocal part that approaches a constant above some occupation number $m_{\rm max}$, we are able to analytically derive the static characteristics of the condensate: the critical density for condensation, the extension of the condensate  for a given size, the shape of the condensate, that is its average occupation number as a function of spatial position, the fluctuations of the condensate and therefore also the mass distribution, that is the probability for finding $m$ unit masses at a given site of the ring.
The derivation is not restricted to the fluid phase for which the description in terms of a grand-canonical partition function holds up to the critical mass density. In our approach the condensed phase becomes analytically accessible due to the assumption that the partition function factorizes into a part $Z_c$ that describes the contribution of the condensate, and a critical background $Z_b$, for which the uniform particle distribution has just the critical density, so that all excess mass is absorbed by the condensate. The factor $Z_c$ can be explicitly calculated when it is independent of the ultralocal part of the weight, i.e., it is of $p(m)$. This happens in particular for the choice of weight functions as it was used in~\cite{evans1}. Otherwise we use what we call the fixed-envelope approximation. Here we approximate $Z_c$ via the probability $P(W)$ of having a condensate extended over $W$ sites with unknown but fixed envelope $h(t)$ for which we use a form that is inspired by numerical simulations of the condensation. The condensate's extension follows in these cases from the maximum of the probability $P(W)$.

As it turns out, within the same fixed-envelope approximation scheme,  when local and ultralocal weights are ``short-ranged", we can tune the extension to scale with the system size as $N^\alpha$ with $0\le\alpha\le 1/2$. When both parts of the weights are long-ranged, the condensate gets localized to a single site as in the ZRP. For long-range local part and short-range ultralocal part, the condensate takes a rectangular form, its height scales proportionally to the system size, while its extension remains constant: features that remind to finite-size scaling of a first-order phase transition.
Therefore hopping rates, leading to PFSS, do not necessarily lead to extended condensates, and if they do so, the shape of the condensate is non-universal.

The paper is organized as follows: In the next section we introduce the model and derive its stationary state from the corresponding master equation. In section \ref{sec3} we discuss short-range interactions with an ultralocal part that is constant above some threshold $m_{max}$. Under mild assumptions on the factorization of the partition function we derive
the critical density for condensation, the scaling of the extension of the condensate, the condensate's shape and its fluctuations, and the single-site distribution of particles, in particular also in the condensed phase. In sections \ref{fixenv} and \ref{sec4} we use a more heuristic argumentation to predict the shape of the condensate for all other choices of weight functions, in particular for power-like decay of local weights. Section \ref{outlook} contains the summary and conclusions. Some supplementary material is provided in the Appendices.

\section{The model and its stationary states\label{sec2}}
We assume that we have $M$ particles placed initially at random on $N$ sites forming a closed chain. Each site $i$ carries a certain number $m_i$ of particles, which may range from zero to $M$. The periodicity implies that $m_{N+1}\equiv m_1$. With rate $u(m_i|m_{i-1},m_{i+1})$ the particle jumps out of site $i$ it occupies, and moves to one of the two neighboring sites. The hopping may be asymmetric: with probability $r$ the particle departures to the right, and with probability $1-r$ to the left. In a computer simulation, this corresponds to choosing at each time step the departure site at random and, if it is not empty, moving a single particle with probability proportional to $u(m_i|m_{i-1},m_{i+1})$ (after a proper normalization) to one of the neighboring sites.
Furthermore, we assume that
\bq
	u(m_i|m_{i-1},m_{i+1}) \equiv f(m_i,m_{i-1})f(m_i,m_{i+1}), \label{urate}
\eq
with some non-negative function $f(m,n)$. The hopping rate is thus a product of two factors, one for each nearest neighbour.
As we shall see, this is the crucial assumption that leads to the pair-factorized steady state. The model is very similar to the one from \cite{evans1}, the only difference is that there the hopping rate was a more general product of two factors $f_1(m_i,m_{i-1})$ and $f_2(m_i,m_{i+1})$. Some aspects of the model were analyzed in \cite{bw1}. In what follows we shall discuss this choice more thoroughly and study some special cases.

At each time $t$, the state of the system is specified by the set of occupation numbers $\vec{m}=\{m_1,...,m_N\}$. Let us denote by $P(\vec{m},t)$ the probability of having the system in a particular state $\vec{m}$ at time $t$. The evolution of $P(\vec{m},t)$ is governed by the master equation:
\bq
	\frac{dP(\vec{m},t)}{dt} = \sum_{\vec{m}'} \left[ W(\vec{m}' \to \vec{m}) P(\vec{m}',t) - W(\vec{m} \to \vec{m}') P(\vec{m},t) \right],
	\label{master}
\eq
where $W(\vec{m}' \to \vec{m})$ is the transition rate from state $\vec{m}'$ to state $\vec{m}$. In this paper we are interested in the steady-state solution of~(\ref{master}), that is in time-independent probabilities $P(\vec{m})\equiv P(\vec{m},t\to\infty)$. As follows from~(\ref{master}),
\bq
	\sum_{\vec{m}'}  P(\vec{m}) W(\vec{m} \to \vec{m}') = \sum_{\vec{m}'}  P(\vec{m}') W(\vec{m}' \to \vec{m}).
	\label{master2}
\eq
If both sides were equal separately for each  $\vec{m}'$, this equation would simply correspond to the detailed balance condition and hence to the equilibrium case. We know, however, that for $r\neq 1/2$ the system cannot be at equilibrium, because there is a net current of particles to the right ($r>1/2$) or to the left ($r<1/2$). It turns out, however, that the solution of (\ref{master2}) does not depend on the hopping asymmetry $r$, and takes the same pair-factorized form either out of or in equilibrium:
\bq
	P(\vec{m}) = \prod_{i=1}^N g(m_i,m_{i+1}) \delta_{m_1+\cdots+m_N,M},
	\label{pfss}
\eq
provided that $g(m,n)$ is some symmetric function $g(m,n)=g(n,m)$ and
\bq
	f(m,n) = \frac{g(m-1,n)}{g(m,n)}. \label{fmn}
\eq
In what follows, we shall skip the Kronecker delta function $\delta_{m_1+\cdots+m_N,M}$ for brevity.
To prove (\ref{pfss}), let us observe that the transition probability $W(\vec{m} \to \vec{m}')$ is non zero and reads $u(m_i|m_{i-1},m_{i+1})$ only if $\vec{m}'=\{\dots,m_{i-1}-1,m_i+1,\dots\}$ or $\vec{m}'=\{\dots,m_i+1,m_{i+1}-1,\dots\}$, i.e., there is one more particle in the configuration $\vec{m}'$ on an arbitrary site $i$ and one less on one of its neighbors. From (\ref{master2}) we obtain:
\ba
	\sum_{i} u(m_i|m_{i-1},m_{i+1}) P(\vec{m}) \nonumber \\
	= \sum_{i} \left[ r u(m_i+1|m_{i+1}-1,m_{i-1}) P(\dots,m_i+1,m_{i+1}-1,\dots) \right. \nonumber \\
	+ \left. (1-r) u(m_{i+1}+1|m_{i}-1,m_{i+2}) P(\dots,m_i-1,m_{i+1}+1,\dots) \right].
	\label{master3}
\ea
We proceed by inserting the guessed pair-factorized steady state (\ref{pfss}), and
\bq
	u(m_i|m_{i-1},m_{i+1}) = \frac{g(m_i-1,m_{i-1})}{g(m_i,m_{i-1})}\frac{g(m_i-1,m_{i+1})}{g(m_i,m_{i+1})}, \label{rate2}
\eq
as follows from (\ref{urate}) and (\ref{fmn}), into equation (\ref{master3}). After canceling some terms we obtain
\ba
	\sum_i g(m_{i-1},m_i-1)g(m_i-1,m_{i+1})g(m_{i+1},m_{i+2}) R_i  \\
	=r \sum_i g(m_{i-1},m_i)g(m_i,m_{i+1}-1)g(m_{i+1}-1,m_{i+2}) R_i \\
	+ (1-r) \sum_i g(m_{i-1},m_i-1)g(m_i-1,m_{i+1})g(m_{i+1},m_{i+2}) R_i,
\ea
where we used the symmetry of $g(m,n)=g(n,m)$ and $R_i=\prod_{j\neq \{i-1,i,i+1\}} g(m_j,m_{j+1})$.
Both terms without the factor $r$ cancel immediately. The remaining terms proportional to $r$ cancel after shifting the index $i\to i+1$ in the last sum. We have therefore proved that when the hopping rate takes the particular form (\ref{rate2}), the system possesses the steady state factorized over pairs of neighboring sites. The proof can be extended to an arbitrary graph \cite{paper3}.

From the above discussion one sees that the hopping rate may be defined by the two-point weight function $g(m,n)$.
It is this very weight function that we shall vary throughout the following sections. In general we shall factorize it into a local interaction factor $K(\vert m-n \vert)$, and an ultralocal factor $p(m)$ (playing the role of an on-site potential) according to
\bq
	g(m,n) = K(|m-n|) \sqrt{p(m)p(n)}, \label{gkpp}
\eq
where both $K(x)$ and $p(m)$ are some positive functions of $x$ and $m$, respectively.
When $K(x)=1$, $g(m,n)$ factorizes and we recover the ZRP with the weight $p(m)$ since every $\sqrt{p(m)}$ appears twice in the product over sites in (\ref{pfss}). As indicated in the introduction, we call it short-range interactions in occupation-number space when $K(\vert m-n\vert)$ or $p(m)$ decay exponentially, and long-range interactions when they decay with some power of the argument.

The most intriguing phenomenon in the ZRP is certainly condensation when a finite fraction of all particles occupies a single site.
Condensation is triggered by the on-site potential $p(m)$, which acts only on particles being at the same site. When we consider in the following PFSS with weights $K(x)$ depending on the difference of occupation numbers $x\equiv\vert m_i-m_{i+1}\vert$ at neighboring sites $i$ and $i+1$, we can explore the effect of nearest-neighbour interactions on the very condensation. Therefore, in what follows
we shall determine the critical density for condensation and the static properties of the condensate for various choices of $K(x)$ and $p(m)$.

\section{Short-range interactions\label{sec3}}
In the first part of this section we will assume the following weights:
\bq
	K(x)=e^{-Jx}, \qquad p(m)=e^{ U\delta_{m,0}}, \label{kpevans}
\eq
with parameters $J$ and $U$, proposed in~\cite{evans1}. The steady state reads:
\bq
	P(m_1,\dots,m_N) = \exp\left( -J\sum_i |m_i-m_{i+1}| + U \sum_i \delta_{m_i,0}\right). \label{pmevans}
\eq
The delta function, reflecting the conservation of particles, was dropped for brevity.
This will be our flag example, although we discuss later a more general case for which $K(x)$ is an arbitrary function that falls off faster than any
power law, and $p(m)$ is constant for $m$ greater than some $m_{\rm max}$. From~(\ref{pmevans}) we see how to interpret $J$ and $U$. Firstly, because $|m_i-m_{i+1}|$ measures the rate of change of $m_i$ with $i$, fast-changing profiles are suppressed the more, the larger $J$ is. Thus, $J$ gives a kind of surface stiffness, which tends to flatten out the profile. 
Secondly, the term $\sum_i \delta_{m_i,0}$ measures the total number of free sites. The larger $U$ is, the more probable are configurations with many empty sites. An interesting observation follows from comparison to the solid-on-solid (SOS) model \cite{sos}. In the SOS model without pinning potential, the probability of a microstate is essentially given by~(\ref{pmevans}) with $U=0$. Such a model does not have a phase transition in 1D. We will see below that it is the insertion of the $U$-term that leads to a liquid-condensed phase transition, even in the 1D equilibrium system.

\subsection{Critical density for condensation\label{sec112}}
In order to study the properties of the steady state we can ignore the fact that the system is out-of-equilibrium, and treat it with conventional methods of statistical mechanics. Thinking of $P(\vec{m})$ as a probability of a microstate, we define the canonical partition function:
\bq
	Z(N,M) = \sum_{\{m_i\}} \prod_i g(m_i,m_{i+1}) \delta_{\sum_i m_i,M} \label{zcanon}
\eq
as well as the grand-canonical one
\bq
	Z_N(z) = \sum_M Z(N,M) z^M = \sum_{\{m_i\}} z^{\sum_i m_i} \prod_i g(m_i,m_{i+1}),
	\label{zgrand}
\eq
which is just the discrete analog of a Laplace transform of $Z(N,M)$, and the fugacity $z$ is determined from
\bq
	\rho =\frac{1}{N}\left<\sum_i m_i\right> = \frac{z}{N}\frac{\partial\ln Z_N(z)}{\partial z} . \label{rhoz}
\eq
The partition function $Z_N(z)$ grows monotonously with $z$, and so does its derivative. Thus the left-hand side of (\ref{rhoz}) grows also with $z$. If the radius of convergence of $Z_N(z)$ is infinite, then for any finite $\rho$ there exists some $z>0$ which obeys~(\ref{rhoz}). Both ensembles, the canonical and the grand-canonical one, are then in our case equivalent in the thermodynamic limit.
We will use this fact to calculate the distribution of particles $\pi(m)$:
\bq
	\pi(m) = \frac{1}{N}\left<\sum_i \delta_{m,m_i} \right>, \label{pimdef}
\eq
that is the probability of having $m$ particles at a randomly chosen site. Taking the average in the grand-canonical ensemble we obtain
\ba
	\pi(m) &=& \frac{1}{Z_N(z)} \sum_{m_2,\dots,m_N} T_{m m_2} T_{m_2 m_3} \cdots T_{m_N m}, \\
	Z_N(z) &=& \sum_{m_1,\dots,m_N} T_{m_1 m_2} T_{m_2 m_3} \cdots T_{m_N m_1} = \mbox{Tr}\, T(z)^N, \label{zcrit}
\ea
where
\bq
	T_{mn}=z^{(m+n)/2}g(m,n).
\eq
If we define now $\phi_m$ to be a normalized eigenvector of $T_{mn}$ to the largest eigenvalue $\lambda_{\rm max}$,
\bq
	\sum_n T_{mn} \phi_n = \lambda_{\rm max} \phi_m ,\label{eq:eigen}
\eq
we obtain for large $N$ that $Z_N(z) \cong \lambda_{\rm max}^N$ and $\pi(m) = \phi_m^2$.
The eigenvector $\phi_m$ has to decay with $m$. Otherwise, $\rho$ calculated from~(\ref{rhoz}) would be infinite in the thermodynamic limit. By analogy to the ZRP we can thus say that the system is in the liquid state --- there is no condensation.

On the other hand, if $Z_N(z)$ has some finite radius of convergence $z_c$, the derivative in (\ref{rhoz}) can either grow to infinity for $z\to z_c$, or tend to some constant. In the first case we have again no condensation, because for any $\rho$ there is some real $z<z_c$ which obeys~(\ref{rhoz}). In the second case, 
there exists a critical density
\bq
	\rho_c = \sum_m m \phi_m^2, \label{rhoc}
\eq
with $\phi_m$ being now the eigenvector for $z=z_c$, above which the grand-canonical ensemble does not exist. This in turn indicates a phase transition from the liquid to the condensed state.

In order to calculate the critical density, it is convenient to assume that $z_c=1$. If $z_c\neq 1$, we can always shift it to one by rescaling the weight $g(m,n)$. If $\phi_m$ and $\lambda_{\rm max}$ are now the eigenvector and the maximal eigenvalue of $T_{mn}(z=1)=g(m,n)$, respectively,
the recipe for obtaining the critical density is i) to find the eigenvector of $g(m,n)$ to the largest eigenvalue, ii) to square its elements, iii) to find its mean value treating $\phi_m^2$ as probabilities. This can be done by numerical diagonalization, 
truncating the matrix to a finite size and keeping track of the size dependence.

In some cases we can find the eigenvector analytically or, at least, we can decide whether $\rho_c$ is finite or not.
First, let us state that because  $g(m,n)\geq 0$ for all $m,n$, from the Frobenius-Perron theorem all entries $\phi_m$ must be non-negative. Moreover, the eigenvector to the largest eigenvalue is not degenerated.
Since $g(m,n)=g(n,m)$, all eigenvalues are real, and eigenvectors to different eigenvalues are orthogonal. This means that all other eigenvectors must have some entries negative, thus if we find a semi-positive vector that obeys the equation:
\bq
	\sum_n g(m,n) \phi_n = \lambda_{\rm max} \phi_m, \label{g_eigen}
\eq
with some constant $\lambda_{\rm max}$, it has to be the one to the largest eigenvalue. 

If $g(m,n)=g(M-m,M-n)$, it follows that $\phi_m=\phi_{M-m}$ and hence $\rho_c=M/2\propto N$ tends to infinity for $N\to\infty$. This is the case of the SOS model, where $g(m,n)=e^{-J|m-n|}$; there is no condensation, regardless of how big the density $\rho$ is. We see the importance of the $U$-term in~(\ref{pmevans}). Its role is to break the symmetry in $g(m,n)$ under the change $(m,n)\to (M-m,M-n)$. When the symmetry is explicitly broken in the occupation number space, from~(\ref{rhoc}) it follows that $\phi_m$ must decay faster than $\sim m^{-1}$ if the critical density has to be finite.

With the above remarks in mind, let us sketch the derivation of $\rho_c$ for the weight (\ref{kpevans}). First, $z_c=1$, so we do not need to shift the weights. As suggested by the functional form of $g(m,n)$, we assume that the eigenvector $\phi_m \propto e^{A\delta_{m,0}+Bm}$ with some constants $A,B$. Indeed, by inserting it into~(\ref{g_eigen}) with $g(m,n)=e^{-J|m-n|+U(\delta_{m,0}+\delta_{n,0})/2}$, one finds that the eigenvalue equation is fulfilled as long as
\ba
	A &=& U/2, \\
	B &=& -J - \ln(1-e^{-U}).
\ea
One sees that when $J<J_0$, where $J_0=U-\ln(e^U-1)$, the eigenvector grows with $m$, therefore the critical density is infinite.
This means that condensation is possible only for $J>J_0$. The unnormalized eigenvector reads $\phi_m \propto e^{U\delta_{m,0}/2-(J-J_0)m}$, and hence the critical density becomes
\bq
	\rho_c=\frac{\sum_{m=0}^\infty m \phi_m^2}{\sum_{m=0}^\infty \phi_m^2} =  \frac{e^{J_0}-1}{(e^{J_0}-e^{-2(J-J_0)})(e^{2(J-J_0)}-1)}. \label{rhoc_evans}
\eq
When $J\to J_0$, the above formula reduces to $\rho_c\approx 1/(e^{2(J-J_0)}-1)$, the result obtained in~\cite{evans1} where the asymptotic form $\phi_m\sim e^{Bm}$ was assumed. 
As a byproduct we obtain the largest eigenvalue:
\bq
	\lambda_{\rm max} = e^{U} + \frac{1}{e^{2J} (1-e^{-U}) -1}. \label{lmax}
\eq

\subsection{\label{ext1}Extension of the condensate\label{sec113}}
\begin{figure}
	\psfrag{mi1}{$m_i$} \psfrag{mi2}{$\left<m_i\right>$} \psfrag{i}{$i$}
	\includegraphics[width=15cm]{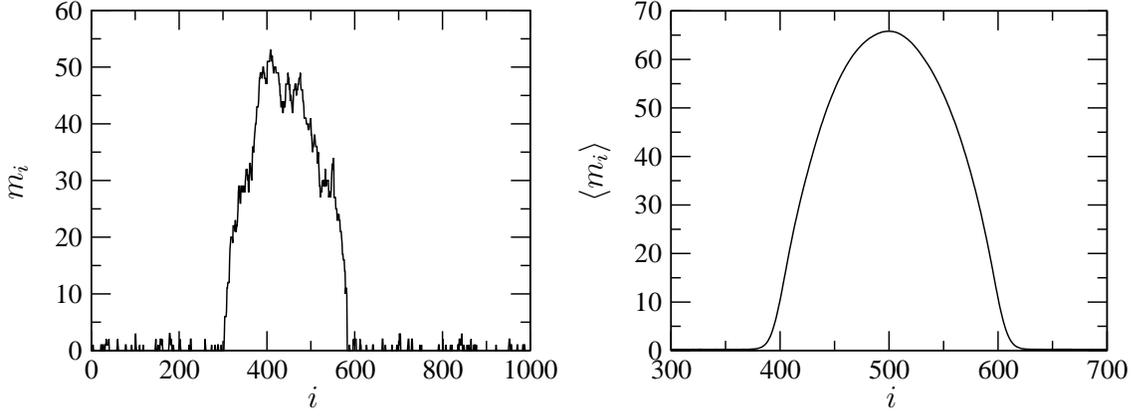}
	\caption{\label{snapshot}Left: a snapshot of the condensate for $N=1000$ and $\rho=10$ obtained in a MC simulation. Right: the shape of the condensate averaged over many realizations, each centered at $i=500$.}
\end{figure}

When the density $\rho$ exceeds $\rho_c$, the condensate emerges in the system.
In figure \ref{snapshot} we present a snapshot of the condensate obtained from numerical simulations for $N=1000$ sites, $J=U=1$ and $\rho=10$. The critical density is $\rho_c=0.2397$, thus the system is deeply in the condensed phase. One sees that the condensate extends over many sites, in contrast to the zero-range process where it is always localized. In~\cite{evans1} the extension $W$, that is the number of sites occupied by the condensate, was estimated to be proportional to $\sqrt{N}$ for sufficiently large systems. The argument was based on a similarity between the envelope of the condensate, and a trajectory of a random walker. Here we will employ another approach, which will allow us to determine not only the extension, but also the average shape of the condensate, its fluctuations and the distribution of particles $\pi(m)$ in the condensed phase.

Let us assume that $\rho>\rho_c$ and consider the weight $P_n$ of a configuration, where $n$ sites share the surplus of particles $M'= N(\rho-\rho_c)$ and where $N-n$ sites form a ``background'' with occupation numbers of order $\rho_c$.
For $n=1$, namely for only one site carrying the condensate, $P_1$ is approximately
\bq
	P_1 = N Z_{N-1}(1) K^2(M'-\rho_c)p(M'-\rho_c).
\eq
The first factor $N$ stands for $N$ possibilities of choosing the site occupied by the condensate.
The second term accounts for $N-1$ ``background'' sites, it is just the partition function (\ref{zcrit}) for $N-1$ sites.
The third term stands for two ``domain walls'' at the borders of the condensate. The fourth term accounts for the on-site potential.
 Since $p(m)\neq 1$ only for empty sites, the fourth term is simply equal to one. Hence the function $p(m)$ influences only $Z_{N-1}$ via the maximal eigenvalue $\lambda_{\rm max}$.
Using the definition (\ref{kpevans}) of $K(x)$ and neglecting $\rho_c$ since it is small in comparison to $M'\to\infty$, we can write:
\bq
	P_1 = N \exp[(N-1)\ln\lambda_{\rm max} - 2J M'].
\eq
For $n=2$, the condensate may occupy either two neighboring sites, or two condensates exist, separated by some number of background sites. The statistical weights for these two situations are, respectively,
\ba	
	P_{2,\,\rm single} &\approx & N Z_{N-2}(1) K(xM')K(xM'-(1-x)M')K((1-x)M'), \\
	P_{2,\,\rm double} &\approx & N^2 Z_{N-2}(1) K^2(xM') K^2((1-x)M'), 
\ea
where $x$ is the fraction of particles occupying the first site, and as before we dropped $\rho_c$ in the arguments. These formulas can be rewritten as
\ba
	P_{2,\,\rm single} &=& P_1 e^{ - \ln\lambda_{\rm max} + J M' (1-|2x-1|)}, \\
	P_{2,\,\rm double} &=& P_1 e^{  - \ln\lambda_{\rm max} +\ln N},
\ea
so that for any $0<x<1$, the two-site single condensate with probability $P_{2,\,\rm single}$ dominates in the thermodynamic limit over i) the one-site condensate $P_1$ and ii) the case $P_{2,\,\rm double}$ when the two sites carrying the condensate are separated.

Similarly, one can show that for $n=3,4,\dots$, $P_n$ grows with $n$. The first observation is that the probability is higher when all $n$ sites keep together. For such a case, assuming that the $i$th site takes a portion $x_i$ of the condensate, we obtain
\bq
	P_n \propto \exp\left( -n \ln\lambda_{\rm max} - J M' \left(\sum_{i=1}^{n-1} |x_{i+1}-x_i| + x_1 + x_n\right) \right), \label{eq:pn}
\eq
where the factor independent of $n$ is dropped, and $x_1+\dots+x_N=1$. The weight $P_n$ has a maximum if
\bq
	f(\vec{x}) = \sum_{i=1}^{n-1} |x_{i+1}-x_i| + x_1 + x_n
\eq
takes the minimum value. To find this minimum, let us observe that $f(\vec{x})$ can be always lowered if we order $x_1,\dots,x_n$ so that $x_1$ and $x_n$ are minimal, and both sequences $x_1,x_2,\dots$ and $x_{n},x_{n-1},\dots$ are increasing. Then, $x_i$ has a maximum at some $i=i_0$. We have:
\bq
	f(\vec{x}) = x_1 + \sum_{i=1}^{i_0-1} (x_{i+1} - x_i)  + \sum_{i=i_0}^{n-1} (x_i - x_{i+1}) + x_n = 2x_{i_0},
\eq
and, since all $x_i$'s must sum up to one, the smallest value is $x_{i_0}=1/n$. Therefore, $P_n \sim \exp(-n \ln\lambda_{\rm max}-2J M'/n)$ clearly grows with $n$ as long as $n\ll \sqrt{N}$ and hence we see that the condensate must be spread over many sites.
One could argue that we did not take into account the entropic contribution, namely on how many ways we can choose the set of $x_1,\dots,x_n$.
But the entropic contribution cannot be larger than $\sim N^{n} = e^{n \ln N}$, so as long as $n$ stays finite, it only changes the constant multiplying $n$ in~(\ref{eq:pn}).
But if $n$ grows with $N$ as $\sim \sqrt{N}$, the two terms $\sim n$ and $\sim M'/n$ become comparable. 
This could suggest that if they both were negative, $P_n$ would take its maximum at $n\sim \sqrt{N}$ and the spatial extension of the condensate would be of order $\sim\sqrt{N}$.
However, the ``naive'' entropic term $n\ln N$ dominates over the $-n \ln\lambda_{\rm max}$ term for very large $N$ and therefore configurations with growing $n$ seem to be more probable. In other words, this argument would imply that the spatial extension should be $\sim N$, and there would be no condensation at all. The problem is caused by the fact that we overestimated the entropic contribution. In the next section we will show that a more accurate calculation yields that the entropy $\sim n$ and the extension is indeed $\sim \sqrt{N}$.

\subsection{\label{shape}Shape of the condensate}

The interesting feature of the condensation discussed so far is that the condensate is extended over many sites.
Apart from the question concerning the scaling of the extension with the system size, to which we will return soon, another interesting question concerns the shape of the condensate. By the shape we understand here the condensate's envelope, that is we ask about the average occupation numbers $\left<m_i\right>$ as a function of the index $i$. A numerical recipe for measuring the envelope in a Monte Carlo (MC) simulation is straightforward: one has to accumulate the shape over many realizations, each time shifting the condensation peak to a common origin. The only difficulty is to find the center of the condensate. This can be done in many ways which are equivalent in the thermodynamic limit. For any finite size, however, we found the following method to be the best. First, one searches for the maximal occupation number, let it be the one at site $i_{\rm max}$. Then, starting from $i_{\rm max}$, one searches for the left $i_{\rm left}$ and the right $i_{\rm right}$ border of the condensate where the occupation numbers drop to the background level $\rho_c$.
The center of the condensate is then defined as $i_0 = (i_{\rm left}+i_{\rm right})/2$. This method preserves sharp borders of the condensate seen in figure \ref{snapshot}, left, quite well, as can be seen in the same figure, right, where we show the envelope obtained as an average over $20000$ MC samples. Since steady-state properties of the model do not depend on the fact whether it is in equilibrium or not, to simulate the system we treated it as if it were in equilibrium with the probability of a microstate (\ref{pmevans}). The model can then be simulated using the standard MC method by picking up a random particle and moving it to a randomly chosen site, accepting or rejecting the move with Metropolis probability. This leads to different dynamic properties than the original model has, but considerably speeds up the convergence towards the steady state in the condensed phase.

A careful inspection of figure \ref{snapshot} shows that the envelope is neither circular (which would be true if particles behaved as in a droplet of water due to some surface tension) nor Gaussian (the most frequent distribution so that it may be naively expected to be observed), but has some parabolic-like shape. A simple argument shows that the shape cannot be determined from ``energetic'' considerations alone, which do not take the entropy into account. Imagine that the condensate has a single maximum and falls monotonically on both sides. Then the ``energy'' which is the logarithm of the steady-state probability (\ref{pmevans}) and reads approximately $-W\ln\lambda_{\rm max}-J\sum_i |m_{i+1}-m_i| = -W\ln\lambda_{\rm max}-2JH$, depends only on the height $H$ and the width $W$, but not on the shape. This means that the principle of energy  minimization alone does not allow us to estimate the shape.

Therefore, to find the envelope of the condensate, one has to take the average over all possible shapes, weighted by $P(\vec{m})$.
Below we will show how to do this. Since the whole reasoning is valid for more general weights (\ref{gkpp}), as long as $K(x)$ falls off faster than any power law, and $p(m)=1$ for $m>m_{\rm max}$, we will keep the discussion quite general. At the end, we will present formulas for the special case (\ref{kpevans}).

Let us remark first that since the grand-canonical partition function (\ref{zgrand}) does not exist in the condensed state, a naive approach via evaluating (\ref{zgrand}) for the fugacity $z$ determined by~(\ref{rhoz}) does not work. One should in principle work with the canonical ensemble, but direct calculation of (\ref{zcanon}) is very hard. This is because in~(\ref{zcanon}) we have $p(m)$ which depends on occupation numbers, as well as $K(|m-n|)$ which depends on their differences. This prevents us from decoupling terms for different sites and to perform the sum over $\{m_i\}$ directly.
One can observe, however, that in the condensed state the system may be split into two parts: a condensate with $W$ sites, and a background with $N-W$ sites. The background can be treated as being at the critical point, because the condensate absorbs all the surplus of balls. Summing over all configurations in the background we shall get the grand-canonical partition function for the system with $N-W$ sites and $\rho_c$ balls per site on average, which reads $Z_{N-W}(1) = \lambda_{\rm max}^{N-W}$, as follows from~(\ref{zcrit}).
From the formulas
\ba
	\left<m_i\right> = \left.\frac{z}{NZ_N(z)} \frac{\partial Z_N(z)}{\partial z}\right|_{z=1}, \\
	\left<m_i(m_i-1)\right> = \left.\frac{z^2}{NZ_N(z)} \frac{\partial^2 Z_N(z)}{\partial z^2}\right|_{z=1},
\ea
applied to the background we obtain that the variance $\left<m_i^2\right>-\left<m_i\right>^2$ is constant.
Thus the total number of particles in the background fluctuates as $\sim \sqrt{N-W}$. This in turn means that $M'$, the number of particles in the condensate, also fluctuates as $\sim \sqrt{N-W}\ll M'$. We shall therefore neglect fluctuations of the condensate's mass, and assume that it takes always $M'$ particles. The width and the height of the condensate can still fluctuate. Moreover, the condensate, from its definition, has occupation numbers growing with $N$. Since we assumed $p(m)=1$ for large $m$, we can drop $p(m)$ in the condensate. On the borders, however, $m_i\approx \rho_c$ and since this is very small in comparison with other $m_i$'s, we can set it to zero.
The statistical weight of the condensate extended to $W$ sites is then
\bq
	Z_{\rm c}(W) \cong \sum_{\{m_k\}} \prod_{k=1}^{W+1} K(|m_k-m_{k-1}|)\delta\left[\sum_{k=1}^W m_k - M' \right], \label{z1}
\eq
where we assumed $m_0=m_{W+1}=0$ at the borders. Here the function $\delta[n]$ denotes the Kronecker delta $\delta_{n,0}$ and will be used from now on for notational convenience instead of $\delta_{n,0}$.
The total weight of having a system composed of the condensate and the background will be assumed as a product of probabilities:
\bq
	P(W) \cong Z_{N-W}(1) Z_{\rm c}(W) = \exp\left((N-W)\ln \lambda_{\rm max} +\ln Z_{\rm c}(W)\right).  \label{pw}
\eq
To find the extension $W$, one has to find the maximum of~(\ref{pw}) with respect to $W$.

We will focus now on calculating $Z_{\rm c}(W)$ defined in~(\ref{z1}). Due to the Kronecker delta constraint and the factors $K(|m_k-m_{k-1}|)$ which suppress large differences in neighboring occupation numbers, we expect that the majority of occupation numbers will be much greater than zero. Therefore, we assume that the summation over $\{m_i\}$ can be extended to negative values without changing $Z_{\rm c}(W)$ too much. We will show later by calculating average occupation numbers and their variances that this assumption is fully justified, that is in fact almost all $m_k>0$ in most probable configurations. We can now rewrite $Z_{\rm c}(W)$ as
\ba
	Z_{\rm c}(W) \approx &\sum_{d_1=-\infty}^\infty \cdots \sum_{d_{W+1}=-\infty}^\infty
	\delta\left[-\sum_{k=1}^{W+1} kd_k - M' \right] \delta\left[\sum_{k=1}^{W+1} d_k \right] \nonumber \\
	&\times \prod_{k=1}^{W+1} K(|d_k|), \label{z2}
\ea
where $d_k=m_k-m_{k-1}$. The first Kronecker delta in the above formula gives the conservation of particles. The second delta reflects fixed boundary conditions according to $m_0=m_{W+1}=0$ which gives $d_1+\dots+d_{W+1}=0$.
Let us introduce an auxiliary function
\ba
	G(W,\vec{u}) = &\sum_{d_1=-\infty}^\infty \cdots \sum_{d_{W}=-\infty}^\infty \delta\left[-\sum_{k=1}^{W} kd_k - M' \right] \delta\left[\sum_{k=1}^{W} d_k \right] \nonumber \\
	&\times \prod_{k=1}^{W} K(|d_k|) e^{d_k u_k}, \label{gen}
\ea
where $\vec{u}=\{u_1,\dots,u_W\}$ are auxiliary variables. We will use them later, let us now only observe that $Z_{\rm c}(W)=G(W+1,\vec{0})$ and hence $G(W,\vec{u})$ is a generating function for the moments of $d_k$:
\bq
	\left<d_k^n\right> = \left[ G(W,\vec{u})^{-1} \frac{{\rm d}^n}{{\rm d}u_k^n} G(W,\vec{u}) \right]_{\vec{u}=0}. \label{eq:dkn}
\eq
Replacing both delta functions in~(\ref{gen}) by their integral representations:
\bq
	\delta[x] = \int_{-i\pi+\epsilon}^{i\pi+\epsilon} \frac{{\rm d}z}{2\pi i} e^{xz},
\eq
and performing the sum over $\{d_k\}$, we obtain:
\bq
	G(W,\vec{u}) =\int_{-i\pi+\epsilon_1}^{i\pi+\epsilon_1} \frac{{\rm d}z}{2\pi i} \int_{-i\pi+\epsilon_2}^{i\pi+\epsilon_2} \frac{{\rm d}v}{2\pi i} e^{F(z,v,\vec{u})},
\eq
where
\bq
	F(z,v,\vec{u}) = -M' z + \sum_{k=1}^W \ln \tilde{K}(u_k+v-k z). \label{fzvu}
\eq
The function $\tilde{K}(x)$ is defined as
\bq
	\tilde{K}(x) = \sum_{d=-\infty}^\infty K(|d|) e^{dx}.
	\label{ktilde}
\eq
The function $G(W,\vec{u})$ can be evaluated in the saddle point: $G(W,\vec{u})\sim e^{F(z,v,\vec{u})}$, with $z=z(\vec{u}), v=v(\vec{u})$ being solutions to the saddle-point equation $\partial_z F(z,v,\vec{u}) = \partial_v F(z,v,\vec{u})=0$. Let us assume for a moment that we have solved this equation and have determined both $z(\vec{u})$ and $v(\vec{u})$. 
With help of~(\ref{eq:dkn}), we can then write:
\ba
	\left<d_k\right> \cong  \left[\frac{{\rm d}}{{\rm d}u_k} F(z(\vec{u}),v(\vec{u}),\vec{u})\right]_{\vec{u}=0} ,\\
	{\rm var}(d_k) = \left<d_k^2\right> - \left<d_k\right>^2 \cong \left[\frac{{\rm d^2}}{{\rm d}u_k^2} F(z(\vec{u}),v(\vec{u}),\vec{u}) \right]_{\vec{u}=0}
	\label{d2}, \\
	{\rm cov}(d_j,d_k) = \left<d_j d_k\right> - \left<d_j\right>\left<d_k\right> \cong \left[\frac{{\rm d}}{{\rm d}u_j} \frac{{\rm d}}{{\rm d}u_k} F(z(\vec{u}),v(\vec{u}),\vec{u}) \right]_{\vec{u}=0}, \label{djdk}
\ea
and similarly for higher-order correlation functions. We stress that in general all derivatives have to be taken for $z(\vec{u}),v(\vec{u})$ being functions of $\vec{u}$, and only at the end one can set $\vec{u}=\vec{0}$.
But for the first moment we have
\bq
	\left<d_k\right> = \left[\pf{F}{z}{1}\pf{z}{u_k}{1} +  \pf{F}{v}{1}\pf{v}{u_k}{1} + \pf{F}{u_k}{1}\right]_{\vec{u}=0},
\eq
where the partial derivatives of $F$ are taken in the saddle point and thus are zero, except of the last one. We thus obtain:
\bq
	\left<d_k\right> = \frac{\tilde{K}'(v-kz)}{\tilde{K}(v-kz)}. \label{dkav}
\eq
The first observation that eliminates $z$ without the need of solving the saddle-point equation is that the averaged peak must be symmetric around its center. This is true even if the system is not in equilibrium, that is when a current of particles flows through the system for $r\neq 1/2$. The symmetry of the shape implies that $\left<m_k\right> = \left<m_{W-k}\right>$, thus $\left< d_k\right> = -\left< d_{W-k}\right>$ and hence $z = (2/W) v$. We obtain:
\bq
	\left<d_k\right> = \frac{\tilde{K}'(v(1-\frac{2k}{W}))}{\tilde{K}(v(1-\frac{2k}{W}))}.
\eq
In addition, because the total number of particles in the condensate equals $M'$, we have for large systems:
\bq
	\sum_{k=0}^W k \left<d_k\right> \cong \int_0^W k \frac{\tilde{K}' \left(v \left(1-\frac{2k}{W}\right)\right)}{\tilde{K}\left(v\left(1-\frac{2k}{W}\right)\right)} {\rm d}k = -M'.
\eq
The last equation, after changing variables, reduces to
\bq
		w = v\left[\frac{1}{2} \int_0^{v}  \frac{x\tilde{K}'(x)}{\tilde{K}(x)} {\rm d} x \right]^{-1/2}, \label{ftox}
\eq
where neither $M'$ nor $W$ appear alone and where we defined the reduced extension $w\equiv W/\sqrt{M'}$. Equation (\ref{ftox}) fixes $w$ as a function of $v$. We can now write
\bq
	\ln Z_{\rm c}(W) \cong F\left(\frac{2v}{w(v)\sqrt{M'}},v,\vec{0}\right), \label{lnzc}
\eq
and combining equations~(\ref{lnzc}), (\ref{fzvu}) and (\ref{pw}) we obtain that the logarithm of the weight of a condensate extended over $W$ sites is given by
\ba
	\ln P(W(v)) = &N\ln\lambda_{\rm max} + \sqrt{M'}\left[-w(v)\ln\lambda_{\rm max} -\frac{2v}{w(v)}\right. \nonumber \\
	& \left.   + \frac{w(v)}{v} \int_0^{v} \ln \tilde{K}(x) {\rm d}x \right]. \label{lnpw}
\ea
Since $W$ does not appear in the above formula explicitly, but only through the relation $W=w(v)\sqrt{M'}$, the extension $W$ is now determined by maximizing~(\ref{lnpw}) with respect to $v$. Hence, $W$ must be proportional to $\sqrt{M'}\sim \sqrt{N}$. Thus, we have proved that the extension grows with the square root of the system size.

One can go one step further and calculate $v$ and $w$, and thus the exact, numeric value of the extension $W$.
The probability (\ref{lnpw}) assumes a maximal value, when its derivative with respect to $v$ is zero. Inserting (\ref{ftox}) into~(\ref{lnpw}) and taking the derivative $d/dv$ we obtain:
\bq
	\left(\ln\lambda_{\rm max} - \ln \tilde{K}(v)\right) \left[\frac{1}{v}-\frac{v}{2}\frac{\tilde{K}'(v)}{\tilde{K}(v)} \frac{1}{\int_0^{v} \frac{x \tilde{K}'(x)}{\tilde{K}(x)}{\rm d}x}
	 \right] =0.
\label{eq:kl}
\eq
One can show (see Appendix A) that the second term in the square brackets is never zero. Only the first term counts and gives:
\bq
	v = \tilde{K}^{-1}(\lambda_{\rm max}). \label{veq}
\eq
Having $v$ and $w$, it is now easy to obtain the shape of the condensate:
\bq
	\left<m_n\right> = \left<\sum_{k=1}^n d_k\right> = \sum_{k=1}^n  \left<d_k\right> ,\label{mndef}
\eq
where $\left<d_k\right>$ is given by~(\ref{dkav}). If we define new variables:
\bq
	h \equiv \left<m_n\right>/\sqrt{M'}, \qquad t = \frac{2n}{w\sqrt{M'}}-1,
\eq
so that $t\in[-1,1]$, we can write the formula for the envelope $h(t)$ as follows:
\bq
	h(t) = \frac{w}{2v} \ln \frac{\tilde{K}(v)}{\tilde{K}(v t)}. \label{mnfinal}
\eq
Thus the shape in reduced variables $t,h(t)$ becomes independent of the system size and the density of particles. The parameters $w,v$ depend on the weights $K(x)$ and $p(m)$ only.

\begin{figure}
	\psfrag{xx}{$t$}\psfrag{yy}{$h(t)$}
	\includegraphics[width=15cm]{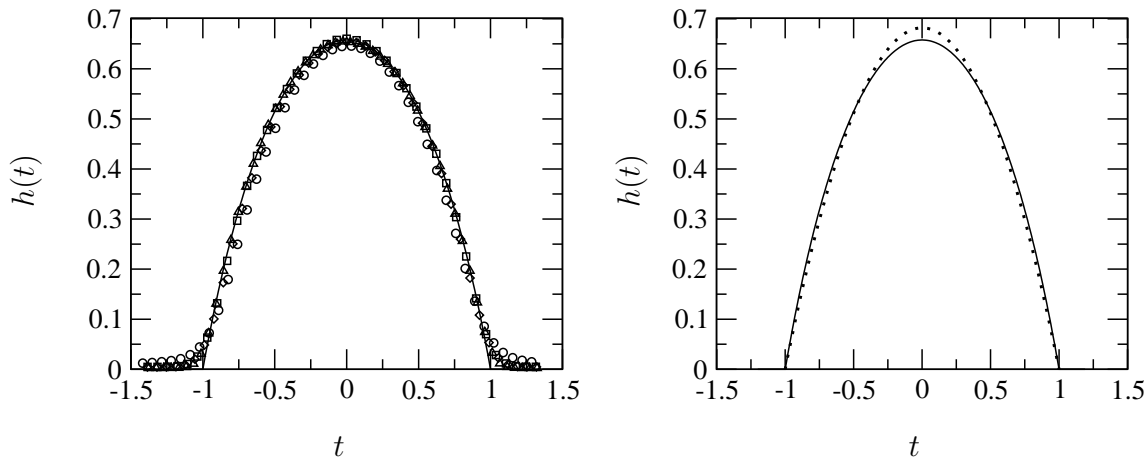}
	\caption{\label{fig_ht}Left: Comparison between the envelope of the condensate from~(\ref{htfinal}) and rescaled MC data for the weights (\ref{kpevans}) with $J=U=1$ for $N=1000$, $\rho=M/N=1$ and $3$ (circles, squares) and $N=4000$, $\rho=1,3$ (diamonds, triangles).
Right: the envelope (\ref{htfinal}) (solid line) compared to the parabola~(\ref{parab}) (dotted line).
}
\end{figure}

Let us now go back to the choice (\ref{kpevans}) of the weight functions.
For this case, one obtains
\bq
	\tilde{K}(x) = \frac{\sinh J}{\cosh J - \cosh x},
\eq
and $v$, calculated from~(\ref{veq}), assumes a very simple form: $v = J-J_0$.
Inserting $v$ obtained from the above equation into~(\ref{ftox}), we can calculate $w$ numerically. The shape is given by~(\ref{mnfinal}) and reads
\bq	
h(t) = \frac{w}{2v} \ln \left( \frac{\cosh J - \cosh v t}{\cosh J - \cosh v}\right),
	\label{htfinal}
\eq
with $w$, $v$ being functions of the parameters $J,U$. For instance, for $J=U=1$ one obtains $v=0.5413$ and $w=2.2005$.
In figure \ref{fig_ht} we show a comparison between~(\ref{htfinal}) and MC simulations for different system sizes. One sees that the curves obtained for different $N,\rho$ and plotted in the rescaled variables $t,h$ agree with the theoretical line. Small deviations in tails can be attributed to finite-size effects and will be discussed in section 3.4, but one sees already in figure \ref{fig_ht} that slopes of the envelope become more steep for increasing system size $N$.

One may ask now the following question. The shape of $h(t)$ is determined by~(\ref{z2}), which is just the partition function for a Brownian excursion: a 1D random walk with steps drawn from $K(x)$ starts at $m=0$ and ends also at $m=0$ after $W$ steps. The only difference in comparison to ordinary excursions is that the area under the trajectory is fixed to some value $M'$. If this area is not fixed, it fluctuates around $\sim W^{3/2}$, the result known for a ``free'' excursion \cite{exc}. Assuming $M'\propto W^{3/2}$ in our formulas we have $w = W/\sqrt{M'} \propto W^{1/4}$ and from (\ref{ftox}) we obtain that
\bq
	v=\frac{6\tilde{K}(0)}{\tilde{K}''(0) w^2} \to 0.
\eq
Thus, by applying Taylor expansion to~(\ref{mnfinal}) one obtains a universal curve, independent of $K(x)$:
\bq
 h(t)=(6/4w)(1-t^2). \label{parab}
\eq
It must be stressed, however, that in our case $M'$ scales always as $W^2$, so the above argument does not hold, and the envelope is not universal. In figure \ref{fig_ht}, right, we see that despite a lack of universality in general, the exact form (\ref{htfinal}) does resemble the parabola from (\ref{parab}) very much, at least in a certain range of $J,U$.

\subsection{Fluctuations\label{sec115}}
In this section we shall consider only the special choice (\ref{kpevans}).
Assuming that the width $W$ of the condensate is fixed, the fluctuations around the average envelope are measured by:
\ba
	{\rm var}(m_n)_W &= \sum_{k=1}^n \sum_{j=1}^n (\left<d_k d_j \right> - \left<d_k\right> \left<d_j\right>) \nonumber \\
	&= \sum_{k=1}^n {\rm var}(d_k) + \sum_{k=1}^n\sum_{j\neq k} {\rm cov}(d_k,d_j), \label{varmn}
\ea
where the subscript $W$ tells us that $W$ is assumed to be constant.
We need to calculate
\ba
	{\rm var}(d_k) = &\pf{F}{u_k}{2} + \pf{F}{z}{2}\left(\pf{z}{u_k}{1}\right)^2 + \pf{F}{v}{2}\left(\pf{v}{u_k}{1}\right)^2 \nonumber \\
	&+ 2 \frac{\partial^2 F}{\partial z \partial u_k} \pf{z}{u_k}{1} + 2\frac{\partial^2 F}{\partial v \partial u_k} \pf{v}{u_k}{1}
	+ 2\frac{\partial^2 F}{\partial z \partial v} \pf{z}{u_k}{1}\pf{v}{u_k}{1}
	\label{vardk}
\ea
and
\ba
	{\rm cov}(d_k,d_j) = &\pf{F}{z}{2}\pf{z}{u_k}{1}\pf{z}{u_j}{1} + \pf{F}{v}{2}\pf{v}{u_k}{1}\pf{v}{u_j}{1}
	+\frac{\partial^2 F}{\partial z \partial v}\left(\pf{v}{u_k}{1}\pf{z}{u_j}{1}+\pf{v}{u_j}{1}\pf{z}{u_k}{1}\right) \nonumber \\
	&+ \frac{\partial^2 F}{\partial z \partial u_k}\pf{z}{u_j}{1} + \frac{\partial^2 F}{\partial z \partial u_j}\pf{z}{u_k}{1} \nonumber \\
	&+ \frac{\partial^2 F}{\partial v \partial u_k}\pf{v}{u_j}{1}+ \frac{\partial^2 F}{\partial v \partial u_j}\pf{v}{u_k}{1},
	\label{covkj}
\ea
where all partial derivatives of $F(z,v,\vec{u})$ defined in~(\ref{fzvu}) are taken in the saddle point with $\vec{u}=0$.
Other derivatives which in principle may appear in the above formula are zero and have been dropped. Since this is rather technical, we postpone the derivation to Appendix B. Here we shall only present the final result:
\bq
	{\rm var}(m_n)_W = \frac{w\sqrt{M'}}{2}\sigma^2\left(\frac{2n}{w\sqrt{M'}}-1\right), \label{varfinal}
\eq
where the size-independent variance $\sigma^2(y)$ is given by 
\bq
	\sigma^2(y) = I_0(y) - \frac{I_0^2(y)}{I_0(1)}  - \frac{I_1^2(y)}{I_2(1)}, \label{sigmay}
\eq
with
\bq
	I_m(t) \equiv  \int_{-1}^{t} y^m \frac{\cosh J \cosh (v y) -1}{(\cosh J-\cosh (v y))^2} {\rm d}y
\eq
derived in Appendix B.
The formula (\ref{varfinal}) would give the fluctuations of the envelope if the width $W$ was kept fixed. In reality, the width also fluctuates around the mean $W=w\sqrt{M'}$. To take this into account we have to sum over all possible widths, each having the probability $P(W)$ given by~(\ref{lnpw}),
\ba
	{\rm var}(m_n) &=& \sum_W P(W) \left<m_n^2\right>_W - \left(\sum_W P(W) \left<m_n\right>_W\right)^2 \\
	&=& \sum_W P(W) \left[ {\rm var}(m_n)_W + \left(\left<m_n\right>_W - \left<m_n\right>\right)^2\right].
\ea
Here $\left<m_n\right>_W$ means the envelope for given $W$, while $\left<m_n\right>$ means the envelope averaged over $W$.
Since the probability $P(W)$ is concentrated around $W=w\sqrt{M'}$ for large systems, the latter can be approximated by $\sqrt{M'} h(t)$, where $t=2n/(w\sqrt{M'})-1$. Also the variance ${\rm var}(m_n)_W$ summed with weights $P(W)$ is well approximated by $\frac{1}{2}w\sqrt{M'} \sigma^2(t)$. Then, from the above equation it follows that
\ba
	{\rm var}(m(t)) \approx \sqrt{M'}&\left[ \frac{w}{2}\sigma^2(t) + \sqrt{M'} \int P(W)\left(h\left(t\frac{w\sqrt{M'}}{W}\right)\frac{w\sqrt{M'}}{W} \right.\right. \nonumber \\
	&\left.\left.-h(t)\right)^2 {\rm d}W \right], \label{avvar1}
\ea
so there is an additional contribution to the variance from fluctuations of the condensate's width.
The probability $P(W)$ can be well approximated by a Gaussian distribution with the mean $w\sqrt{M'}$ and variance ${\rm var}(W) \equiv \sqrt{M'} s^2$ with
\bq
	s^2 = w \frac{\frac{w^2}{4} \tilde{K}'(v) -\tilde{K}(v)}{v\tilde{K}'(v)}.
\eq
This formula comes from expanding the probability $P(W(v))$ around the minimum for $v=v_0$, and is true for any $\tilde{K}(x)$.
For our special case (\ref{kpevans}) it reads
\bq
	s^2 = w\frac{w^2/4 +(\cosh(v) - \cosh(J))/\sinh(v)}{v}.
\eq
Then, equation (\ref{avvar1}) can be rewritten as
\ba
	{\rm var}(m(t)) \approx \sqrt{M'}&\left[ \frac{w}{2}\sigma^2(t) + \frac{wM'^{3/4}}{\sqrt{2\pi s^2}} \right. \nonumber \\
	&\left.\times\int e^{-\frac{w^2\sqrt{M'}}{2s^2}(x-1)^2}\left(\frac{1}{x}h\left(\frac{t}{x}\right)-h(t)\right)^2 {\rm d}x \right]. \label{avvar}
\ea
\begin{figure}
	\psfrag{xx}{$t$} \psfrag{s2}{$\mbox{var}(m(t))/\sqrt{M'}$}\psfrag{yy}{$\left<h(t)\right>$}
	\includegraphics[width=15cm]{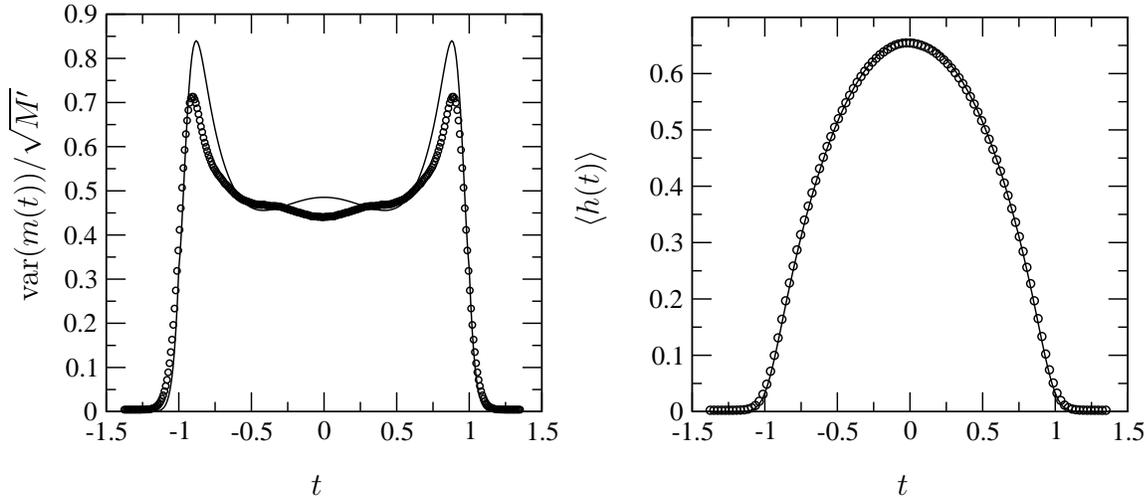}
	\caption{\label{fig_var} Left: The variance ${\rm var}(m_n)$ obtained from MC simulations for $N=4000,\rho=3$ (circles), compared to~(\ref{avvar}) evaluated numerically (thin line). Right: the averaged envelope $\left<h(t)\right>$ from~(\ref{avh(t)}), taking into account small fluctuations of $M'$.
}
\end{figure}

In figure~\ref{fig_var}, left, we compare the result (\ref{avvar}) and the one obtained by means of MC simulations.
To a good approximation, the variance ${\rm var}(m_n)$ grows linearly with $W\sim \sqrt{M'}$.
This means that in the thermodynamic limit the number of occupation numbers being less than zero becomes negligible, because typically $m_n = \left<m_n\right> \pm \sqrt{{\rm var}(m_n)} \sim W \pm \sqrt{W}$, so that extending the summation in (\ref{z2}) to negative values was fully allowed because it leads to self-consistent results.
At the center of the condensate, for $t=0$, the formula (\ref{avvar}) simplifies to
\bq
	{\rm var}(m(0)) \approx \sqrt{M'}\left( \frac{w}{2}\sigma^2(0) + \frac{s^2}{w^2} h^2(0) \right). \label{var0}
\eq
In the next section we shall see that this formula is very important for the problem of the distribution of particles.

As a by-product we obtained the variance $\sqrt{M'}s^2$ of the distribution of the width $W$. This allows us to calculate the envelope $h(t)$ averaged over fluctuations of $W$:
\bq
	\left<h(t)\right> \approx \frac{wM'^{1/4}}{\sqrt{2\pi s^2}} \int e^{-\frac{w^2\sqrt{M'}}{2s^2}(x-1)^2}\frac{1}{x}h\left(\frac{t}{x}\right) {\rm d}x. \label{avh(t)}
\eq
In figure \ref{fig_var}, right, we show the curve obtained from the above equation. It fits to the MC data points even better than equation (\ref{htfinal}), especially in the tails. This indicates that smooth tails are mainly due to the fluctuations of the condensate's width.
This effect clearly vanishes with increasing system size, i.e., is a typical finite-size effect.

It turns out that not only $\left<m_n\right>$ and ${\rm var}(m_n)$ but also the distribution of $m_n$ for any fixed $n$ can be predicted.
Since $m_n$ is a sum of many (almost independent) random variables $d_k$'s, each of them having finite variance, the distribution tends to a Gaussian, according to the Central Limit Theorem, provided that $1\ll n \ll W$, that is we are not too close to the borders. In the thermodynamic limit we can safely assume that all $m_n$ have Gaussian distributions.

\subsection{Distribution of particles\label{sec116}}
Let us denote by $\pi_{\rm b}(m), \pi_{\rm c}(m)$ the probability of finding $m$ particles at a background site and a condensate site, respectively. The distribution $\pi_{\rm b}(m)\propto \phi_m^2$, as follows from the partition function (\ref{zgrand}) at the critical point.
In particular, for the previous weights one obtains $\pi_{\rm b}(m)\propto \left(\frac{e^{-J}}{1-e^{-U}}\right)^{2m}e^{U\delta_{m,0}}$.
The distribution in the condensate, $\pi_{\rm c}(m)$, is given by summing Gaussian distributions with mean $\left<m_n\right>$ and variance ${\rm var}(m_n)$ over all $n=1,\dots,W$ condensate sites:
\bq
	\pi_{\rm c}(m) \propto \sum_{n=1}^W \frac{1}{\sqrt{2\pi {\rm var}(m_n)}} \exp\left[-\frac{(m-\left<m_n\right>)^2}{2{\rm var}(m_n)} \right],
\eq
which gives
\bq
	\pi_{\rm c}(m) \propto \int_0^1 \frac{ {\rm d} t}{\sqrt{2\pi{\rm var}(m(t))}}\exp \left[-\frac{(m-\sqrt{M'}h(t))^2}{2{\rm var}(m(t))} \right].
\eq
By changing variables $t\to y=h(t)$ one obtains
\bq
	\pi_{\rm c}(m) \propto \int_0^{h(0)} (h^{-1})'(y) \frac{\exp \left[-\frac{(m-\sqrt{M'}y)^2}{2{\rm var}(m(h^{-1}(y)))} \right]}{\sqrt{2\pi{\rm var}(m(h^{-1}(y)))}}  {\rm d} y.
\eq
Hence the probability $\pi_{\rm c}(m)$ may be viewed as a convolution of $(h^{-1})'(y)$ with a Gaussian distribution which smears the thermodynamic-limit distribution $\pi_{\rm c}(m)\sim (h^{-1})'(m/\sqrt{M'})$. Since the variance grows like $\sqrt{M'}$, the smearing acts on distances $\sim M'^{1/4}$ and influences the profile of $\pi_{\rm c}(m)$ only at $y\approx h(0)$, because $(h^{-1})'(y)$ is  narrow only at $y=h(0)$. We can therefore assume ${\rm var}(m(t)) = {\rm var}(m(0))$ because it affects the distribution only at $t=0$, and write
\bq
	\pi_{\rm c}(m) \propto \int_0^1 \frac{ {\rm d} t}{\sqrt{2\pi{\rm var}(m(0))}}\exp \left[-\frac{(m-\sqrt{M'}h(t))^2}{2{\rm var}(m(0))} \right],
\eq
with ${\rm var}(m(0))$ given by~(\ref{var0}).
Since the main contribution to $\pi_{\rm c}(m)$ comes from the flat region in $h(t)$, the condensate's peak tends to a Gaussian for large systems.

\begin{figure}
	\psfrag{mm}{$m$}\psfrag{pi}{$\pi(m)$}
	\includegraphics[width=15cm]{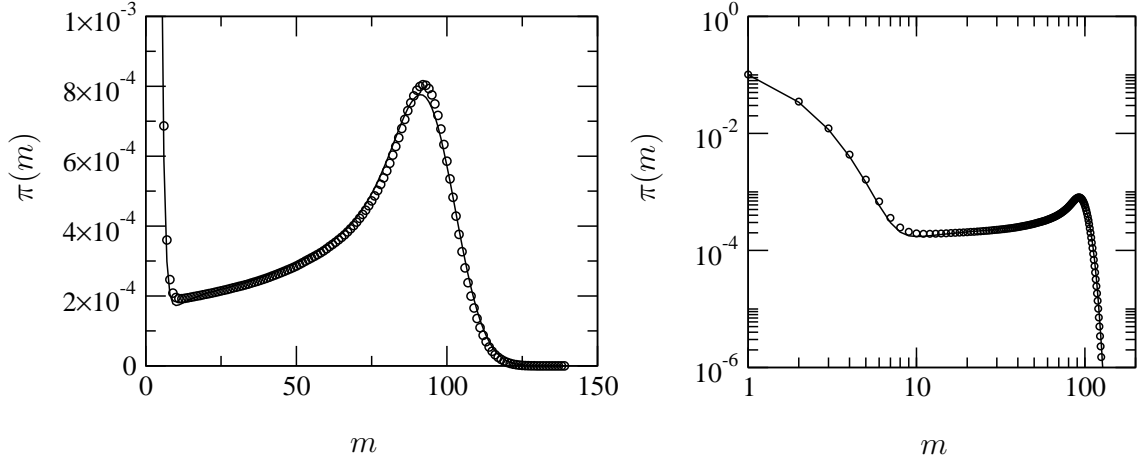}
	\caption{\label{fig:pi}Left: the distribution of particles $\pi(m)$ obtained from~(\ref{pim}) (thin line) and from MC simulations for $N=8000,\rho=3$ (circles). Right: the same on a log-log scale.
}
\end{figure}

The probability $\pi(m)$ of finding $m$ particles at any site is additively composed according to
\bq
\pi(m)=\frac{N-W}{N}\pi_{\rm b}(m)+\frac{W}{N}\pi_{\rm c}(m). \label{pim}
\eq
In figure~\ref{fig:pi} we show a comparison between $\pi(m)$ obtained from~(\ref{pim}) after inserting $\pi_{\rm b}(m)$ and $\pi_{\rm c}(m)$ normalized to probabilities, with MC data. The agreement is very good.

\section{The case when $p(m)\neq 1$ for all $m$\label{fixenv}}
As long as $K(x)$ falls off faster than any power-law, and $p(m)={\rm const}$ for $m$ larger than some
$m_{\rm max}$, the method from the previous section directly applies.
This means that one can calculate the envelope of the condensate, its fluctuations and the distribution of particles.
Since $p(m)$ can be multiplied by any factor without changing ``physical'' quantities, the condition on $p(m)$ can be assumed to be $p(m>m_{\rm max})=1$.

However, if $p(m)$ does not tend to one for large $m$, the partition function cannot be approximated by~(\ref{z1}). An important example is
\bq
	K(x)\sim e^{-a|x|^\beta}, \qquad p(m)\sim e^{-b m^\gamma}, \label{expab}
\eq
for $a,b,\beta,\gamma>0$, with both $K(x)$ and $p(m)$ decaying fast with $x$ and $m$, respectively.
In order to have the condensation, the grand-canonical partition function must have a finite radius of convergence. This means that the function
\bq
	\sum_m g(m,n) z^m = e^{-b n^\gamma} \sum_m e^{ -a|m-n|^\beta -b m^\gamma} z^m,
\eq
must diverge for $z$ larger than some critical $z_c$. It is easy to check that $z_c$ is finite for $\gamma<1$ and there is condensation above some critical density. For $\gamma>1$, the radius of convergence is infinite, so there is no condensation for $\gamma>1$. The value of $\beta$ is not important here and can be arbitrarily large.

Next we have to determine for which values of $\beta,\gamma$ the condensate is extended.
Let us calculate the contribution from the condensate having $\sim N$ particles at a single site. This is the same approach as in section \ref{ext1}:
\bq
	P_1 \approx N c^{N-1} p(N) K^2(N) \approx N c^{N-1} e^{-b N^\gamma -2aN^\beta}.
\eq
As before, $N$ stands for $N$ possible condensate's positions, $c^{N-1}$ with some constant $c$ accounts for background sites, $p(N)$ is the on-site potential at the condensed site and $K^2(N)$ stands for the two domain walls. The contribution from the condensate residing on two adjacent sites, having respectively $N/2+\epsilon$ and $N/2-\epsilon$ particles, reads
\ba
  P_2 &\approx & N c^{N-2} K(N/2+\epsilon) K(N/2-\epsilon) K(2\epsilon) p(N/2+\epsilon) p(N/2-\epsilon)\\
  & \approx & N c^{N-2} e^{-a(N/2+\epsilon)^\beta-a(N/2-\epsilon)^\beta - a|2\epsilon|^\beta - b(N/2+\epsilon)^\gamma-b(N/2-\epsilon)^\gamma }.
\ea
The contribution is large only for $\epsilon\approx 0$. We can then write:
\bq
	\ln P_1/P_2 \approx -\ln c - 2a(1-2^{-\beta})N^\beta -b(1-2^{1-\gamma}) N^\gamma.
\eq
Since $\gamma<1$,  $1-2^{1-\gamma}<0$. On the other hand, $1-2^{-\beta}>0$ and hence
\bq
	\ln P_1/P_2 \simeq - N^\beta + N^\gamma,
\eq
where the proportionality factors have been skipped.
For $\gamma>\beta$ we thus obtain $P_1/P_2 \to \infty$, so the condensate must be located on a single site. For $\gamma<\beta$, however, the contribution from two sites is larger than from a single one. Similarly, one can show that the contribution grows for 3,4,5,... sites. Therefore the condensate must be extended, but we still do not know, for how much.

\subsection{Fixed-envelope approximation}
\begin{figure}
	\begin{center}
	\includegraphics[width=8.5cm]{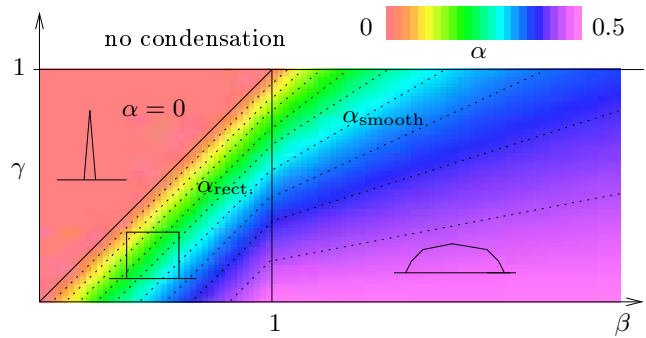}
	\end{center}
	\caption{\label{phase1}Phase diagram for $K(x)\sim e^{-|x|^{\beta}}$ and $p(m)\sim e^{-m^\gamma}$. Values of the exponent $\alpha$ in the extension $W\sim M'^\alpha$, $\alpha_{\rm rect}=(\beta-\gamma)/(\beta-\gamma+1)$ for the rectangular and $\alpha_{\rm smooth}=(\beta-\gamma)/(2\beta-\gamma)$ for the smooth condensate, are represented by the colour (gray) code. Dotted lines mark $\alpha=0.05,0.1,\dots,0.45$.}
\end{figure}

We have learned that fluctuations of the occupation numbers can be neglected if $p(m)\to 1$. This is also the case here, because the fluctuations are determined by the variance of $K(x)$, which is again finite. We shall thus assume that, for sufficiently large systems, the envelope of the condensate is essentially fixed, modulo some small fluctuations around it.
Then, the logarithm of the weight of the condensate extended to $W$ sites is
\ba
	\ln P(W) \approx  &-&W\ln\lambda_{\rm max} + W s + \sum_k \ln K\left(\left<m_{k+1}-m_k\right>\right) \nonumber \\
	&+& \sum_k \ln p\left(\left<m_{k}\right>\right).
\ea
Here $s$ is some ``entropic'' factor due to small fluctuations of $m_k$'s.
Values of $\left<m_k\right>$ are equal to $Hh(2k/W-1)$, where $H=M'/W$ is the height of the condensate and $h(t)$ is some fixed ``envelope'' having the same meaning as in section \ref{shape}, but its functional form may be  unknown. The differential term $\left<m_{k+1}-m_k\right>$ is more difficult. It can behave in two distinct ways. First, if we assume that $h(t)$ is a ``smooth'' (differentiable) function, the difference can be rewritten as $\frac{2M'}{W^2}h'(t)$ with $t=2k/W-1$. A second possibility is that $h(t)$ has a rectangular shape, $h'(t)$ does not exists at $t=\pm 1$. Thus, we may have in principle two extended condensates: the one whose shape resembles that in the previous sections, with smooth $h(t)$, which we will call ``smooth'', and a ``rectangular'' one where $h(t)$ looks like a step function in the limit of large $N$. We can write
\ba
  \ln P_{\rm smooth}(W) \approx & W&\left[ c + \int_0^1 \ln K\left(\frac{2M'}{W^2}h'(t)\right) {\rm d}t \right. \nonumber \\
  &+& \left.  \int_0^1 \ln p(Hh(t)){\rm d}t \right], \label{pnormgen} 
\ea
\bq
  \ln P_{\rm rect}(W) \approx  W c + 2\ln K\left(h(1)\frac{M'}{W}\right) + W \int_0^1 \ln p(Hh(t)){\rm d}t ,\label{prectgen}
\eq
where $c=s-\ln\lambda_{\rm max}$ is some unknown constant.
To find the extension $W$, we must find the larger of the two maximum values of the $P(W)$'s.
Assuming now the weights (\ref{expab}), with $0<\gamma<1$, $\beta>\gamma$, we have:
\ba
	\ln P_{\rm smooth}(W) \approx & W&\left[ c - a\left(\frac{2M'}{W^2}\right)^\beta \int_0^1 |h'(t)|^\beta {\rm d}t \right. \nonumber \\
	&-& \left. b\left(\frac{M'}{W}\right)^\gamma \int_0^1 |h(t)|^\gamma {\rm d}t\right], 
\ea
\bq
	\ln P_{rect}(W) \approx  W c - a\left(\frac{M'}{W}\right)^\beta 2|h(1)|^\beta - bW\left(\frac{M'}{W}\right)^\gamma \int_0^1 |h(t)|^\gamma {\rm d}t.
\eq
The integrals are just some constants. The value $c$ must be smaller than zero, because otherwise $P(W)$ has no maximum which is in contradiction to the fact that the condensate is extended. To estimate the extension, we can drop constants (remembering that they are all positive, except $c$) and write
\ba
 \ln P_{\rm smooth}(W) &\sim &  -W - W\left(\frac{M'}{W^2}\right)^\beta  - W\left(\frac{M'}{W}\right)^\gamma, \label{pnormal} \\
 \ln P_{\rm rect}(W) &\sim & -W - \left(\frac{M'}{W}\right)^\beta - W\left(\frac{M'}{W}\right)^\gamma. \label{prect}
\ea
Let us consider $P_{\rm smooth}(W)$ first. If $\beta>1/2$, it has a maximum for $d\ln P_{\rm smooth}/dW = 0$ which reads
\bq
	(1-2\beta)\left(\frac{M'}{W^2}\right)^\beta  + \left(\frac{M'}{W}\right)^\gamma = -1,
\eq
and hence
\bq
	W\sim M'^{(\beta-\gamma)/(2\beta-\gamma)},
\eq
because the exponents of $M'$ in both terms must be equal. We also obtain the maximal value:
\bq
	\ln P_{\rm smooth}(W) \sim - M'^{(\gamma\beta +\beta - \gamma)/(2\beta-\gamma)}.
\eq
If $\beta<1/2$, the above equation has no solution, and $\ln P_{\rm smooth}(W)$ takes its maximal value at $W\sim 1$, which would mean ``delta-like". This is impossible, so this solution is obviously not valid for $\beta<1/2$, where the condensate must therefore be rectangular and described by~(\ref{prect}), which has the maximum for
\bq
	\beta M'^\beta W^{-\beta-1} - (1-\gamma) M'^\gamma W^{-\gamma} = 1.
\eq
From the above equation we obtain the extension of the rectangular condensate
\bq
	W\sim M'^{(\beta-\gamma)/(\beta-\gamma+1)},
\eq
as well as
\bq
	\ln P_{\rm rect}(W) \sim - M'^{\beta/(\beta-\gamma+1)}.
\eq
If $\beta>1/2$, both equations (\ref{pnormal}) and (\ref{prect}) are valid. But for $M'\to\infty$, it turns out that $\ln P_{\rm rect}(W)>\ln P_{\rm smooth}(W)$ for $1/2<\beta<1$. Therefore, collecting everything together we see that the rectangular-shaped condensate dominates for $\beta<1$, and the smooth condensate dominates for $\beta>1$. In figure~\ref{phase1} we show a phase diagram which summarizes all the above results. One sees there that the extension $W\sim M'^\alpha$ can be tuned to any $0\leq \alpha<1/2$, but cannot be larger than $1/2$.

\section{The case when $K(x)$ has heavy tails\label{sec4}}
In this section we consider $K(x)$ decaying as a power-law: $K(x)\sim x^{-\nu}$. For this case, the analytic method from Section 3 cannot be applied even if $p(m)$ is constant above some $m_{\rm max}$. The reason is that the series (\ref{ktilde}) does not converge for $K(x)$ decaying more slowly than exponentially. On the other hand, the assumption that fluctuations vanish in the thermodynamical limit, a crucial step in the fixed-envelope approximation, is also not well justified a priori, because $K(x)$ may have infinite variance.
Fortunately, it will turn out to be quite easy to show that the condensate is never extended over more than some finite number of sites, so there will be no need to use any of the two methods mentioned.

\subsection{Double power law: $K(x)\sim x^{-\nu}$ and $p(m)\sim m^{-b}$\label{221}}
Let us first discuss the criterion for condensation. For $b>0$, the eigenvector $\phi_m$ has the following form for large $m$:
\bq
	\phi_m \sim m^{-b/2-\nu}, \label{phi2}
\eq
which can be checked by inserting it into~(\ref{g_eigen}). The critical density
\bq
	\rho_c \sim \sum_m m^{-b-2\nu+1}
\eq
is finite when $\nu>1-b/2$. The above formula is valid also for $\nu<0$, that is when configurations with large differences between neighboring occupation numbers are favored. For $\nu=0$, that is when $K(x)\sim$ const, we recover the ZRP case: the critical density is finite for $b>2$.
If $b<0$, the solution (\ref{phi2}) is not valid, but then $\phi_m$ has to grow with $m$. The critical density is therefore infinite for $b<0$.

To estimate the extension, let us first calculate the contribution from the condensate having $\sim N$ particles at a single site. Following the same method as before one obtains
\bq
	P_1 \approx N c^{N-1} p(N) K^2(N) \approx c^{N-1} N^{1-b-2\nu}.
\eq
The contribution from the condensate residing on two adjacent sites, taking respectively $x$ and $1-x$ fractions of particles, is:
\bq
  P_2 \approx N c^{N-2} p(xN) f\left((1-x)N\right) K(xN) K\left((1-x)N\right) K\left(|1-2x|N\right).
\eq
For fixed $x$, in the limit $N\to\infty$, we obtain
\bq
	P_2 \approx c^{N-2} N^{1-2b-3\nu} (x(1-x))^{-b-\nu} |1-2x|^{-\nu}.
\eq
Clearly $P_2 \sim P_1 N^{-b-\nu}$ and since $-b-\nu<-b/2-1$, it vanishes in comparison to $P_1$ in the thermodynamic limit.
Even if we assume that the difference $\epsilon$ in occupation numbers of both sites is small ($x\approx 1/2$), we have
\bq
	P_2 \approx c^{N-2} N^{1-2b-2\nu} K(\epsilon),
\eq
which is still by a factor $\sim N^{-b}$ smaller than $P_1$. Performing also the sum over $\epsilon$ gives an additional factor not larger than $N^{1-\nu}$. Thus $P_2 \sim P_1 N^{1-b-\nu} < P_1 N^{-b/2}$ and the ratio $P_2/P_1$ tends always to zero. The conclusion is therefore that the condensate, if it exists, is always localized at a single site for any $b,\nu$, precisely as for the ZRP.

This localization is, however, not easy to observe in numerical simulations.
If the system is initially prepared in a state with a uniform distribution of particles and the simulation starts, after even a very long time one usually sees the condensate occupying more than one site. This is caused by a small probability to decay from the extended to a single-site condensate. Once the condensate becomes localized on a few sites, there are large energetic barriers between condensates of decreasing sizes. On the other hand, if one starts the simulation from the state with only one site occupied by all particles, one observes that this state does not decay to an extended condensate in the course of the simulation.

\subsection{Mixed decay law: $K(x)\sim x^{-\nu}$ and $p(m)=e^{U\delta_{m,0}}$\label{222}}
Let us finally briefly discuss the case where $K(x)\sim x^{-\nu}$ follows a power law as in the previous subsection but $p(m)=e^{U\delta_{m,0}}$ is of short-range nature as in (\ref{kpevans}).
Again, we shall first estimate the probabilities $P_1,P_2$ of having the condensate on one or two sites:
\ba
	P_1 \sim N c^{N-1} N^{-2\nu}, \\
	P_2 \sim N c^{N-2} N^{-2\nu} 4^\nu K(\epsilon).
\ea
One sees that both terms depends on $N$ in the same way and none becomes negligible for $N\to\infty$. Also higher $P_n$ share the same feature:
\bq
	P_n \sim N c^{N-n} N^{-2\nu} 4^\nu \left[\sum_\epsilon K(\epsilon)\right]^{n-2}.
\eq
This suggests that the extension is determined by the ratio $\sum_x K(x)/c$ (or more precisely: by the inverse of its logarithm) and is constant if this ratio is smaller than one. If $\sum_x K(x) = \infty$, that is for $\nu\leq 1$, or when $\sum_x K(x)/c>1$, $P_n$ grows with $n\to\infty$ and the condensation is impossible. This means that the prefactor in $K(x)\sim x^{-\nu}$ does matter --- consider for instance $K(x)=1/(1+x/a)^\nu$ which behaves as $a^\nu x^{-\nu}$ for large $x$. When $a$ is small, $\sum_x K(x)$ is large and there is no condensation.

We can get the same prediction by applying naively the fixed-envelope approach, developed in section~\ref{fixenv}, and forgetting about problems caused by (perhaps) large fluctuations of neighboring occupation numbers. We use equations~(\ref{pnormgen}) and (\ref{prectgen}) with $K(x)\sim x^{-\nu}$ and $p(m)=0$, because $p(m)$ vanishes now in the condensate. We find that $\ln P_{\rm smooth} \sim -M^{1/2}$ with $W\sim M^{1/2}$ and $\ln P_{\rm rect} \sim - {\rm const}$ with $W\sim {\rm const}$, so that $\ln P_{\rm smooth}<\ln P_{\rm rect}$ and the condensate is rectangular and of fixed extension.

\section{Conclusions and Outlook\label{outlook}}
We have shown how to tune the shape of the condensate during spontaneous symmetry breaking in mass transport models on a ring topology. The shape of the condensate and the scaling of its extension with the system size are non-universal. They depend on the competition between the ultralocal and local contributions (in occupation number space) to the weight factors whose product over all pairs of sites determines the stationary state. Analytical predictions were possible even for the condensed phase above the critical mass density for which the grand-canonical partition function is no longer convergent; a partition function that factorizes over the condensate and the critical background turns out to be the appropriate approximation scheme, as the excellent agreement with numerical simulations of the shape and the single-site mass distribution have demonstrated.

Some of the results presented in this paper we have already used in~\cite{paper3} for predicting the onset of condensation
in case of anisotropic hopping in two dimensions, where the two-dimensional system could be dimensionally reduced to an effectively one-dimensional ZRP. Moreover, in~\cite{paper3} we generalized the topology on which certain classes of hopping rates lead to PFSS from a one-dimensional ring to arbitrary connected and undirected graphs. From the theoretical point of view it remains challenging to derive the phase structure from these known PFSS in higher dimensions that cannot be reduced to effectively one-dimensional processes.

Vice versa, from experimental observations of the shape and the scaling of the width with the system size, one may trace back the class of hopping interactions that are compatible with the observations. When atoms condense on a crystal surface, they can migrate and build extended islands. As experiments on the  deposition of clusters \cite{exp} or fabrication of quantum dots \cite{qd}
show, the islands can be extended in the direction perpendicular to the surface. Currently it is open as to whether hopping rates leading to PFSS are able to reproduce the shape and the typical size of the islands of atoms obtained in such experiments.

\ack{B.W. and W.J. thank the EC-RTN Network "ENRAGE" under grant No.~MRTN-CT-2004-005616 and the Alexander von Humboldt Foundation's grant No.~3.4-Fokoop-DEU/1117877 for support.
B.W. would like to thank the International Center for Transdisciplinary Studies (ICTS) at Jacobs University for its hospitality and support of several visits during this collaboration.
}

\appendix
\section*{Appendix A}\label{appendixa}
\setcounter{section}{1}
We will show that
\bq
	Q\equiv \frac{1}{v}-\frac{v}{2}\frac{\tilde{K}'(v)}{\tilde{K}(v)} \frac{1}{\int_0^{v} \frac{x \tilde{K}'(x)}{\tilde{K}(x)}{\rm d}x},\label{eq:t1}
\eq
entering (\ref{eq:kl}) is never equal to zero. 
From the definition of $\tilde{K}(x)$,
\bq
	\tilde{K}(x) = \sum_{d=-\infty}^\infty K(|d|) e^{dx},
\eq
we see that $\tilde{K}'(x)+\tilde{K}(0)>\tilde{K}(x)$ for $x>0$. This in turn means that $\tilde{K}'(x)/\tilde{K}(x)$ grows with $x$. Therefore, $x \tilde{K}'(x)/\tilde{K}(x)$ is convex for $x>0$. The area under a convex function $f(x)$ over a range $(0,v)$ is smaller than the area of a trapezoid whose parallel edges are placed at $x=0,v$ and range from zero to $f(a),f(b)$, respectively. This leads to the following inequality:
\bq
	\int_0^{v} \frac{x \tilde{K}'(x)}{\tilde{K}(x)}{\rm d}x < \frac{v^2}{2} \frac{\tilde{K}'(v)}{\tilde{K}(v)},
\eq
and finally gives
\bq
	Q < \frac{1}{v} - \frac{v}{2}\frac{\tilde{K}'(v)}{\tilde{K}(v)} \frac{1}{\frac{v^2}{2} \frac{\tilde{K}'(v)}{\tilde{K}(v)}} = 0,
\eq
so that $Q<0$ for all $v>0$.

\appendix
\section*{Appendix B}\label{appendixb}
\setcounter{section}{2}
We shall evaluate the formulae (\ref{vardk}) and (\ref{covkj}) in order to find the variance ${\rm var}(m_n)_W$ from (\ref{varmn}).
They can be simplified if one observes that by differentiating the saddle point equation with respect to $u_i$ one obtains:
\ba
	\pf{F}{z}{2}\pf{z}{u_i}{1} + \frac{\partial^2 F}{\partial z \partial v}\pf{v}{u_i}{1} = - \frac{\partial^2 F}{\partial z \partial u_i}, \label{spd1} \\
		\pf{F}{v}{2}\pf{v}{u_i}{1} + \frac{\partial^2 F}{\partial z \partial v}\pf{z}{u_i}{1} = - \frac{\partial^2 F}{\partial v \partial u_i}.
\label{spd2}
\ea
One can insert equations~(\ref{spd1}), (\ref{spd2}) into~(\ref{vardk}) and (\ref{covkj}), and rewrite~(\ref{varmn}) as
\bq
	{\rm var}(m_n)_W = \sum_{k=1}^n \pf{F}{u_k}{2} + \sum_{k=1}^n\sum_{j=1}^n \left( \frac{\partial^2 F}{\partial z \partial u_k}\pf{z}{u_j}{1}+ \frac{\partial^2 F}{\partial v \partial u_k}\pf{v}{u_j}{1} \right). \label{varmn2}
\eq
This formula is far less complicated, but still needs derivatives of $z,v$ taken at the saddle point with $\vec{u}=0$. Instead of solving the saddle-point equation with $\vec{u}\neq 0$ directly, and differentiating the solution over $u_i$, we can use the following trick: we will calculate the derivatives of $F$ and obtain $\partial z/\partial u_i, \partial v/\partial u_i$ from equations~(\ref{spd1}), (\ref{spd2}). Let us calculate first:
\ba
\pf{F}{u_k}{2} &=& c(k),  \\
	\frac{\partial^2 F}{\partial z \partial u_k} &=& - \frac{k}{z_0} c(k), \\
	\frac{\partial^2 F}{\partial v \partial u_k} &=& \frac{1}{v_0} c(k), \\
	\pf{F}{z}{2} &\cong & \frac{1}{z_0^2} \sum_k k^2 c(k) \cong \frac{1}{z_0^2} \frac{W^3}{8} (I_0(1)+I_2(1)), \\
	\pf{F}{v}{2} &=& \frac{1}{v_0^2} \sum_k c(k) \cong \frac{1}{v_0^2} \frac{W}{2} I_0(1), \\
	\frac{\partial^2 F}{\partial z \partial v} &=& -\frac{1}{z_0 v_0} \sum_k k c(k) \cong -\frac{1}{z_0 v_0} \frac{W^2}{4} I_0(1),
\ea
where we defined
\ba
	c(k) &\equiv & \frac{\cosh J \cosh \left[v\left(1-\frac{2k}{W}\right)\right] -1}{\left[\cosh J-\cosh \left(v\left(1-\frac{2k}{W}\right)\right)\right]^2}, \\
	I_m(t) &\equiv & \int_{-1}^{t} y^m \frac{\cosh J \cosh (v y) -1}{(\cosh J-\cosh (v y))^2} {\rm d}y.
\ea
From~(\ref{spd1}) and (\ref{spd2}) one sees that it must be also $\pf{z}{u_i}{1}\sim c(i)$ and $\pf{v}{u_i}{1}\sim c(i)$. The proportionality factors in these formulas must be functions of type $A+Bi$, because the only dependence on $i$ in equations~(\ref{spd1}), (\ref{spd2}), after dividing by $c(i)$, is linear in $i$. Inserting $\pf{z}{u_i}{1}=(A_z+B_z i) c(i)$ and $\pf{v}{u_i}{1}=(A_v+B_v i) c(i)$ into~(\ref{spd1}) and (\ref{spd2}) and solving for $A_v,A_z,B_v,B_z$, we obtain
\ba
	\pf{z}{u_j}{1} &=& -z \frac{4}{I_2(1) W^2} \left(1- \frac{2j}{W}\right)c(j), \\
	\pf{v}{u_j}{1} &=& v \frac{2}{I_2(1) W} \left(-1-\frac{I_2(1)}{I_0(1)} + \frac{2j}{W}\right)c(j).
\ea
Inserting this into~(\ref{varmn2}), we finally obtain the variance for fixed width $W$:
\bq
	{\rm var}(m_n)_W = \frac{w\sqrt{M'}}{2}\sigma^2\left(\frac{2n}{w\sqrt{M'}}-1\right), 
\eq
where we defined a size-independent variance $\sigma^2$ as follows:
\bq
	\sigma^2(y) = I_0(y) - \frac{I_0^2(y)}{I_0(1)}  - \frac{I_1^2(y)}{I_2(1)}.
\eq
This is precisely equation (\ref{sigmay}).

\end{document}